\documentclass[]{iopart}

\usepackage{iopams}

\usepackage{graphicx}   
\usepackage{dcolumn}    
\usepackage{bm}         

\begin{document}

\title{Structure and elastic properties of Mg(OH)$_2$ from density functional theory}

\author{Pawe\l{} T Jochym,
        Andrzej M Ole\'s,
        Krzysztof Parlinski,
        Jan \L{}a\.{z}ewski,
        Przemys\l{}aw Piekarz and
        Ma\l{}gorzata Sternik}

\address{ Institute of Nuclear Physics, Polish Academy of Sciences,
              Radzikowskiego 152, PL-31342 Krak\'{o}w, Poland }

\ead{Pawel.Jochym@ifj.edu.pl}

\date{\today}

\begin{abstract}
The structure, lattice dynamics and mechanical properties of the
magnesium hydroxide have been investigated with static density
functional theory calculations as well as {\it ab initio\/}
molecular dynamics. The hypothesis of a superstructure existing in
the lattice formed by the hydrogen atoms has been tested. The
elastic constants of the material have been calculated with static
deformations approach and are in fair agreement with the
experimental data. The hydrogen subsystem structure exhibits signs
of disordered behaviour while maintaining correlations between
angular positions of neighbouring atoms. We establish that the
essential angular correlations between hydrogen positions are
maintained to the temperature of at least 150 K and show that they
are well described by a physically motivated probabilistic model.
The rotational degree of freedom appears to be decoupled from the
lattice directions above 30K.
\end{abstract}

\pacs{62.20.dq, 65.40.-b, 71.15.Mb}

\submitto{\JPCM}

\section{Introduction}
\label{sec:intro}

The magnesium hydroxide --- Mg(OH)$_2$ --- is a very simple
mineral, called brucite, often regarded as a good example of a
material with structurally bound OH group. The large number of
known silicate phases with such structurally bound OH units
\cite{Finger89,Kanzaki91} motivates high interest in the role
played by the OH groups and the physical properties induced by
them of this type of crystals in the Earth interior. Brucite was a
subject of numerous experimental
\cite{Kruger89,Duffy91,Redfern92,Fei93,Parise94,Catti95,Duffy95}
and theoretical \cite{Parise94,Sherman91,DArco93,Raugei99} studies
in the past. Despite its simple structure, its physical properties
remain still unexplored
--- in fact even its crystallographic symmetry has not been fully
established \cite{Mookherjee06}, although the question about its
structure has been addressed several times in the past
\cite{Parise94,Sherman91,DArco93,Raugei99}.

\begin{figure}
\begin{center}
\includegraphics[width=0.8\columnwidth]{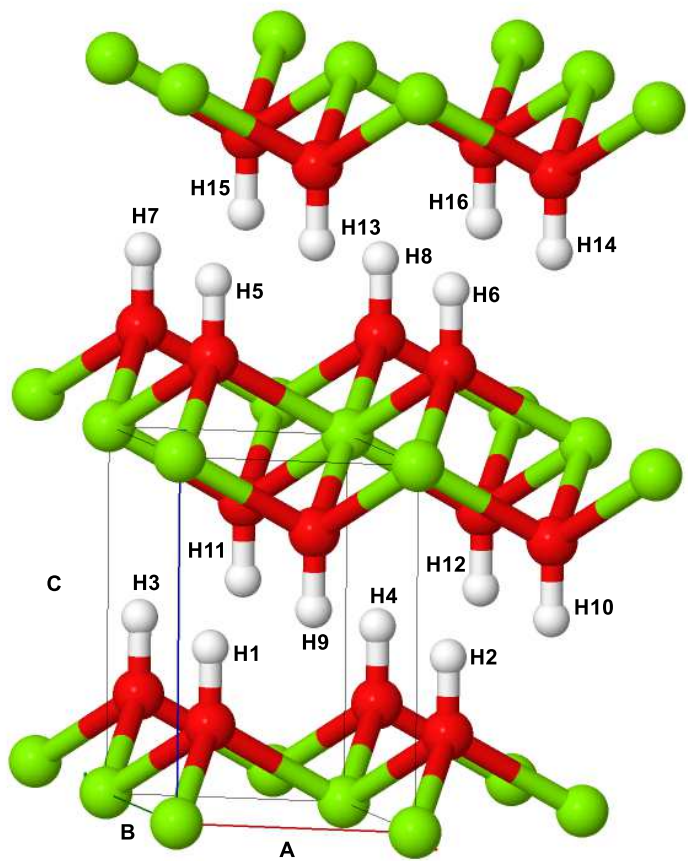}
\end{center}
\caption{\label{fig:structure} Schematic view of the structure of
a brucite crystal. The figure shows $2\times 2\times 2$ supercell
used in the numerical calculations, while the primitive unit cell
is indicated by the (smaller) grey box. Dark grey (red) and light
grey (green) balls show the positions of O and Mg atoms,
respectively. The labels of the hydrogen atoms shown by light
(white) balls are used to analyse their structural correlations in
the text.}
\end{figure}

\begin{figure}
\begin{center}
\includegraphics[width=0.5\columnwidth]{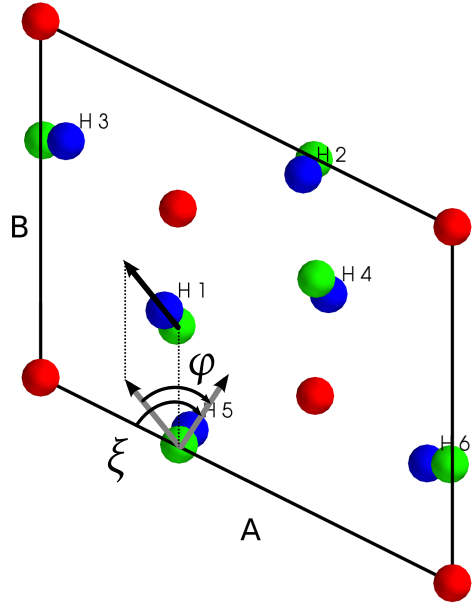}
\end{center}
\caption{\label{fig:structureP-3} Schematic atomic positions in
the $P\overline{3}$ structure of a brucite crystal. The primitive
unit cell is the size of $\sqrt{3}\times\sqrt{3}\times 1$ of the
unit cell of the $P\overline{3}m$ structure shown in
figure~\ref{fig:structure}. The corners are occupied by Mg atoms
(grey, red balls), and each H atom (dark, blue ball) is
accompanied by an O atom (light, green ball). The labels of the
hydrogen atoms are used later in the text to investigate their
spatial correlations. The angles shown here for h5 atom represent
its relative angular coordinate $\varphi$ with respect to the
reference atom H1, and an absolute angular coordinate $\xi$.}
\end{figure}

The crystal lattice of the magnesium hydroxide is a simple,
layered, hexagonal structure with layers of magnesium interlaced
with layers of hydroxyl groups. This basic structure was firmly
established long time ago \cite{Zigan67} as having
$P\overline{3}m1$ symmetry with O-H bond aligned on a three-fold
axis (see figure~\ref{fig:structure}). Further research
\cite{Partin94} showed unexpectedly high thermal displacements of
hydrogen atoms. The models proposed until now
\cite{Parise94,Catti95,Mookherjee06,Megaw73,Parise98,Parise99}
deal with temporary and spatially averaged positions of the
hydrogen subsystem and suggest that the hydrogen atoms are in fact
displaced from the high symmetry points on the three-fold axis to
form a new lattice with $P\overline{3}$ symmetry and averaged 1/3
occupancy of three equivalent $6i$ Wyckoff positions $(x, 2x, z)$,
with either $x>1/3$ (so called XGT arrangement)
\cite{Parise94,Catti95,Mookherjee06,Parise98,Parise99}, or $x<1/3$
(so called XLT arrangement) as proposed by Megaw \cite{Megaw73}.
The schematic placement of atoms in this larger unit cell is
presented in figure \ref{fig:structureP-3}. The case for the XGT
arrangement is further strengthened by the neutron diffraction
experiment of Desgranges {\it et al.\/} \cite{Desgranges96}
indicating that protons are displaced into XGT positions even at
ambient pressure.

All the above models could reflect true structure of the brucite
crystal but due to the performed spatial and temporal averaging of
the structure they may miss the dynamical aspect. In the present
paper we deal with this aspect of the magnesium hydroxide
structure. Firstly, we are going to show results of the static,
elastic constants calculations for the crystal followed by the
results of the density functional theory (DFT) molecular dynamics
(MD) calculations of the brucite crystal's structural properties.
Secondly, we analyse the dynamical aspects of the structure and
demonstrate that the hydrogen atoms are not frozen in their
equilibrium positions but fluctuate strongly, displaying some
unexpected dynamical structure characterized by long range spatial
correlations.

The paper is organized as follows. In section \ref{sec:calc} the
calculation method is presented. It serves to derive elastic
constants of the magnesium hydroxide which are reported and
compared with available experimental values in section
\ref{sec:elast}. Next, in section \ref{sec:corrh}, we investigate
the dynamical model of the hydrogen structure, showing that the
short-range correlations persist and relative hydrogen positions
may be described by a simple probabilistic model presented in
section \ref{sec:model}). The paper is concluded in section
\ref{sec:summa}.

\section{Calculation setup}
\label{sec:calc}

We have performed all calculations of the structure and elastic
properties of the magnesium hydroxide using {\sc
vasp},\cite{Kresse96} a DFT based code employing the generalized
gradient approximation (GGA) \cite{Kresse99} with Perdew, Burke,
and Ernzerhof (PBE) \cite{PBE96} projector augmented-wave (PAW)
pseudopotentials \cite{Blochl94,Kresse99}. The reciprocal space
integration has been carried out over the set of special points
generated according to the Monkhorst-Pack scheme
\cite{Monkhorst76}.

The static calculations have been carried out with a single
hexagonal unit cell shown by a grey box in
figure~\ref{fig:structure}, a $\sqrt{3}\times\sqrt{3}\times1$
supercell (figure~\ref{fig:structureP-3}) and reciprocal space
sampling based on $2\times 2\times 2$ lattice. The calculation of
elastic constants has used the standard finite deformation method
described in detail in the previous publications
\cite{Jochym99,Jochym00}. We have applied displacements between
$\{0.1\% ;\, 0.1^\circ\}$ and $\{1\% ;\, 1^\circ\}$ for the
lattice vector lengths and angles, respectively.

The schematic structure of the crystal in the $2\times 2\times 2$
supercell is presented in figure~\ref{fig:structure} together with
the adopted numbering scheme of hydrogen atoms used later in
section \ref{sec:dynamics} in the analysis of inter-atomic
correlations. The MD calculations have been conducted with both
$\sqrt{3}\times\sqrt{3}\times1$ and $2\times 2\times 2$ unit cells
with the standard energy cut-off, E$_\mathrm{cut}$=400eV,
determined by the used pseudopotentials. We have performed several
3 ps runs with 1-1.5 fs time step for temperatures ranging from 10
K to 150 K in micro-canonical ensemble (constant total energy)
with at least 10 ps thermalization runs carried out with Nos\'e
thermostat technique \cite{Nose84}. The trajectories collected
from the MD runs were further analysed to extract positional
correlations using the computer code developed for this purpose.

To check the influence of the supercell size on obtained results
we have calculated --- using standard direct method
\cite{Parlinski97,Phonon} --- the force constants matrix and full
lattice dynamics for the larger ($3\times 3\times 3$) system as
well as the smaller one used in the MD calculations ($2\times
2\times 2$). The obtained exponential dependence of force
constants on distance shows that the inclusion of the next layer
of unit cells in the system contributes less than 1/1000 of the
overall interaction between atoms in the system. Furthermore, the
dispersion curves and phonon density of states spectra, calculated
for both ($2\times 2\times 2$ and $3\times 3\times 3$) systems,
display no significant difference, which means that the inclusion
of the additional layer of unit cells does not modify the lattice
dynamics of the system in any significant way and the conclusions
concerning lattice vibrations drawn from the present calculations
are robust.

\section{Elastic constants}
\label{sec:elast}

\Table{ \label{tab:cij-hisym} Elastic constants for increasing
pressure $P$ (all quantities in GPa) obtained for the high
symmetry ($P\overline{3}m1$) unit cell of the magnesium
hydroxide.}
\begin{tabular}{*{6}{l}}
\br
 $P$ & $C_{11}$ & $C_{33}$ & $C_{12}$ & $C_{13}$ & $C_{44}$ \\
\mr
1.4 & 126.6 & 534.0 & 112.7 & 39.6 & 61.8 \\
2.1 & 131.2 & 541.8 & 116.1 & 40.8 & 64.3 \\
2.7 & 135.5 & 549.5 & 118.9 & 41.5 & 66.9 \\
\br
\end{tabular}
\endtab

We have completed a series of static, elastic constants
calculations using various sizes and symmetries of the unit cells
and atomic positions: high-symmetry ($P\overline{3}m1$)
experimental unit cell \cite{Zigan67}, low-symmetry
($P\overline{3}$) $\sqrt{3}\times\sqrt{3}\times{}1$ supercell,
both with standard (1\%) and small deformations (0.1\%). In
general, some results presented in table \ref{tab:cij-hisym} for
the high-symmetry unit cell show fairly good agreement with
experimental data \cite{Jiang06}, but others exhibit certain
peculiarities which motivated our further investigation. While the
obtained values of $C_{11}$ are rather close to experimental ones,
the calculated elastic constants $C_{12}$, $C_{13}$ and $C_{44}$
are systematically about twice higher than the experimental values
(table \ref{tab:cij-exper}). The largest difference close to one
order of magnitude was obtained for $C_{33}$ at $P\simeq 1$ GPa.

\Table{ \label{tab:cij-exper} Elastic constants of the magnesium
hydroxide for increasing pressure $P$ (all quantities in GPa),
measured by Jiang {\it et al.\/} \cite{Jiang06} with Brillouin
scattering.} \lineup
\begin{tabular}{*{6}{l}}
\br
 $P$ & $C_{11}$ & $C_{33}$ & $C_{12}$ & $C_{13}$ & $C_{44}$ \\
\mr
 {10$^{-4}$}  & 159.0(16) & \049.5(7) & 43.3(17) & 11.1(25) & 22.8(4) \\
 1.1(1)       & 164.1(10) & \059.2(7) & 45.5(8)  & 12.3(8)  & 25.2(3) \\
 2.1(1)       & 169.3(9)  & \073.2(7) & 46.2(8)  & 16.6(7)  & 28.9(3) \\
 3.5(1)       & 179.6(10) & \094.0(8) & 52.0(9)  & 22.1(8)  & 33.4(3) \\
 4.8(1)       & 184.3(8)  &  110.0(6) & 53.1(6)  & 28.2(5)  & 36.8(3) \\
\br
\end{tabular}
\endtab

During calculations we have observed that the system shows large sensitivity
of elastic constants to external stress applied to the unit cell.
In some cases non-hydrostatic stress of approximately 1-2 GPa induced
twofold change in $C_{33}$ and $C_{13}$ constants.

The discrepancy between theoretical and experimental values of
elastic constants can be partially resolved by careful relaxation
of the internal degrees of freedom of the system in the deformed
state. The standard procedure calls for cell deformation while
keeping the same fractional positions of atoms followed by the
internal degrees of freedom relaxation. Since the hydrogen atoms
start at high symmetry positions on the three-fold axis the
straight-forward minimization procedure keeps this symmetry. Only
after lifting this symmetry constraint, the true ground state
obtained after deformation (with lower total energy) could be
reached. The elastic constants obtained following this procedure
become fairly close to the results of Brillouin scattering
experiment summarized in table~\ref{tab:cij-exper}. As a result of
this relaxation the hydrogen atoms are displaced from their
starting high-symmetry positions $(\frac{1}{3},\frac{2}{3},z)$ and
$(\frac{2}{3},\frac{1}{3},1-z)$.

\Table{ \label{tab:cij-losym} Elastic constants for increasing
pressure $P$ (all quantities in GPa) obtained for the low symmetry
($P\overline{3}$) unit cell of the magnesium hydroxide.} \lineup
\begin{tabular}{*{6}{l}}
 \br
 $P$ & $C_{11}$ & $C_{33}$ & $C_{12}$ & $C_{13}$ & $C_{44}$ \\
 \mr
 0.01 & 130.6 & \048.5  & 70.3 &  10.0 & 20.4 \\
 0.96 & 133.5 & \051.9  & 67.8 & \09.8 & 24.2 \\
 2.0  & 143.4 & \082.6  & 73.4 &  17.7 & 29.3 \\
 3.0  & 150.0 & \093.5  & 76.3 &  15.7 & 32.9 \\
 4.0  & 153.9 &  104.2  & 76.1 &  22.1 & 36.6 \\
 \br
\end{tabular}
\endtab

Prompted by the above anomalies and concepts presented in the
published research on the structure of Mg(OH)$_2$, we have
calculated the same properties for the larger
$\sqrt{3}\times\sqrt{3}\times{}1$ unit cell with $P\overline{3}$
symmetry and hydrogen atoms displaced into XGT positions as
suggested recently by Mookhejee and Stixrude \cite{Mookherjee06}.
This suggestion agrees also with the earlier results in the
literature \cite{Parise94,Catti95,Parise98,Parise99}. Indeed, the
values obtained for this phase and presented in
table~\ref{tab:cij-losym} show much better agreement with
experiment (table~\ref{tab:cij-exper}). We have also observed
considerably weaker sensitivity of calculated values to the
external stress applied to the system.

Moreover, during optimization of the magnesium hydroxide
structure, we have observed that positions of hydrogen atoms
around the three-fold axis are not very well constrained in the
system, i.e., they probably move in a very shallow potential well.
This observation is in concordance with previous experimental
evidence \cite{Partin94}. It suggests that the structure of the
magnesium hydroxide crystal may be more complicated than it is
currently accepted.

\begin{figure}
\begin{center}
\includegraphics[width=0.6\columnwidth]{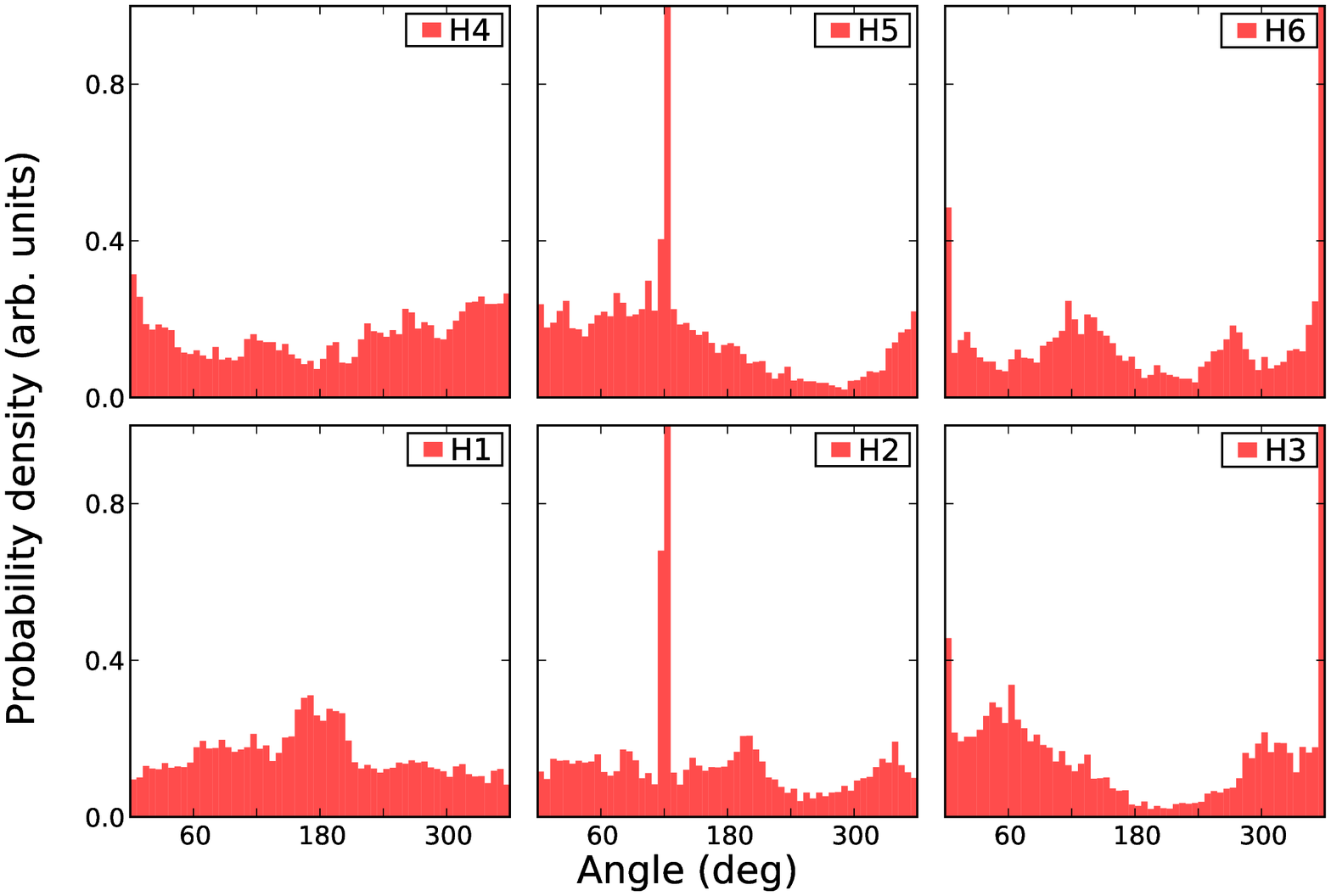}
\end{center}
\begin{center}
\includegraphics[width=0.6\columnwidth]{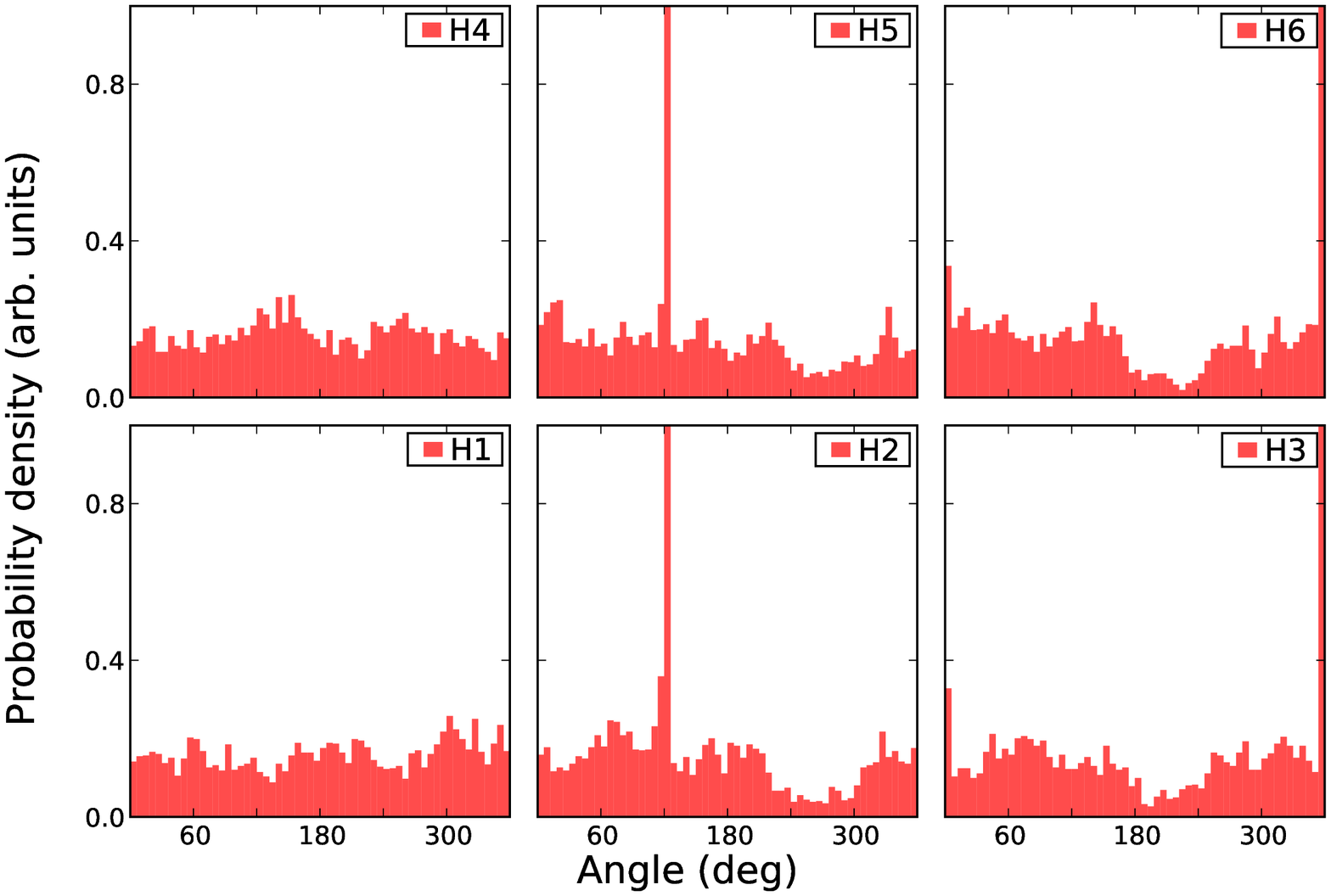}
\end{center}
\caption{\label{fig:corlatt} Absolute angular hydrogen
distribution in P$\overline{3}$ unit cell. Top panels show results
obtained from MD run at 244 K, and bottom ones at 46 K. The maxima
at angle $\varphi\simeq 120^{\circ}$ obtained for H2 (H5) atoms
are 2.05 and 1.2 ( 1.2 and 1.57 ) at high and low temperature,
respectively. Positions of hydrogen atoms with labels
H1,H2,$\cdots$,H6 are shown in figure 1.}
\end{figure}

\section{Dynamical structure}
\label{sec:dynamics}

\subsection{Correlations between hydrogen positions}
\label{sec:corrh}

In order to explore further the issue of dynamical hydrogen
structure in the magnesium hydroxide, we have used the DFT-based
MD technique. The first calculation setup used the
$\sqrt{3}\times\sqrt{3}\times{}1$, P$\overline{3}$ unit cell
proposed by Mookherjee and Stixrude \cite{Mookherjee06}. The data
have been collected from measurement runs of 3 ps length each
preceded by at least 10 ps thermalization run, for several
temperatures between 15 K and 250 K.

The examination of the MD trajectories shows large amplitudes of
hydrogen movements even for runs performed at low temperature.
This result corresponds to large thermal displacements observed in
the experiment \cite{Partin94}. On the one hand, some experimental
results \cite{Desgranges96}, showing clear symmetry of the
hydrogen locations, may be seen as contradicting such observation,
while on the other hand, correlations between the hydrogen
movements could explain this discrepancy.

\begin{figure}
\begin{center}
\includegraphics[width=0.6\columnwidth]{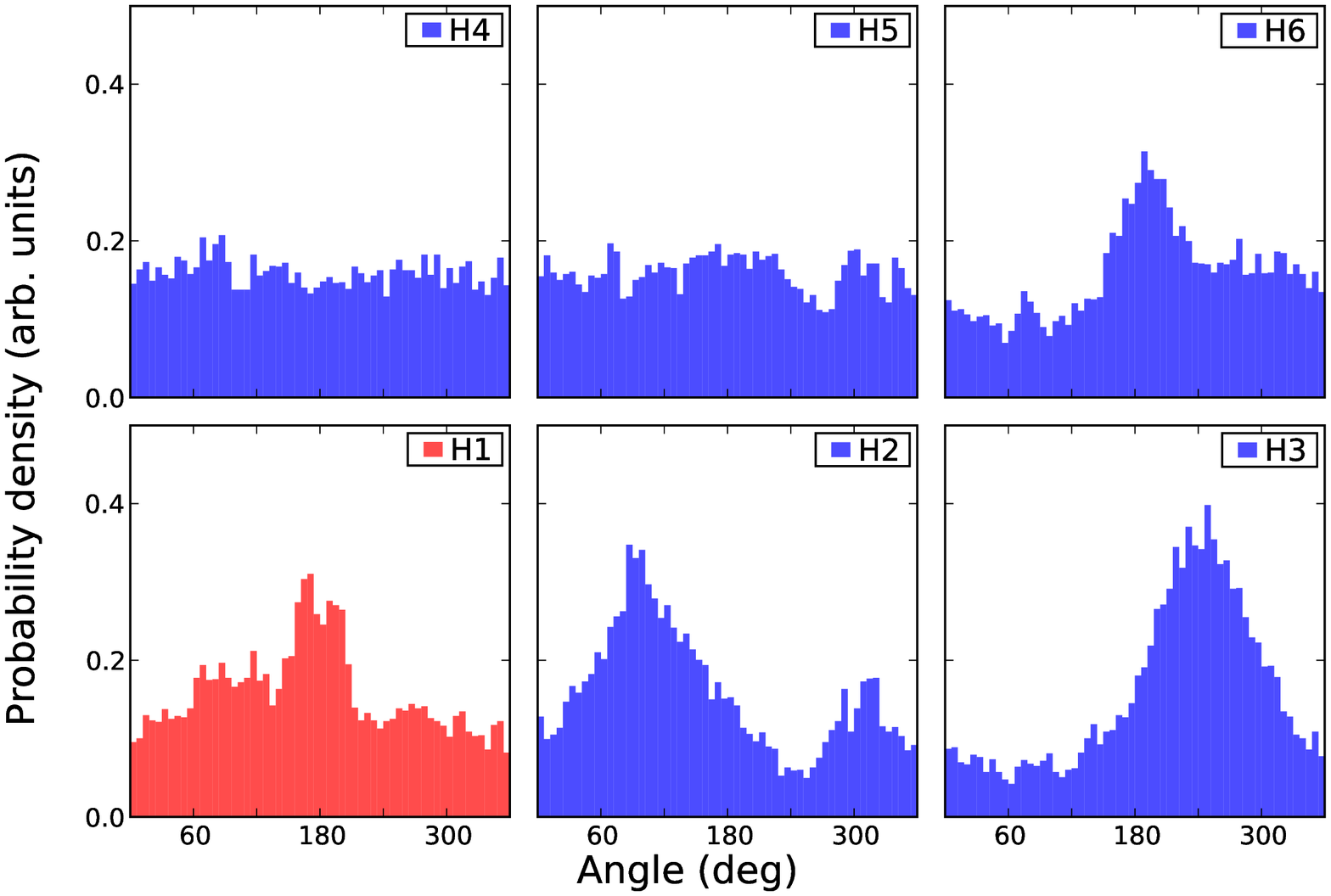}
\end{center}
\begin{center}
\includegraphics[width=0.6\columnwidth]{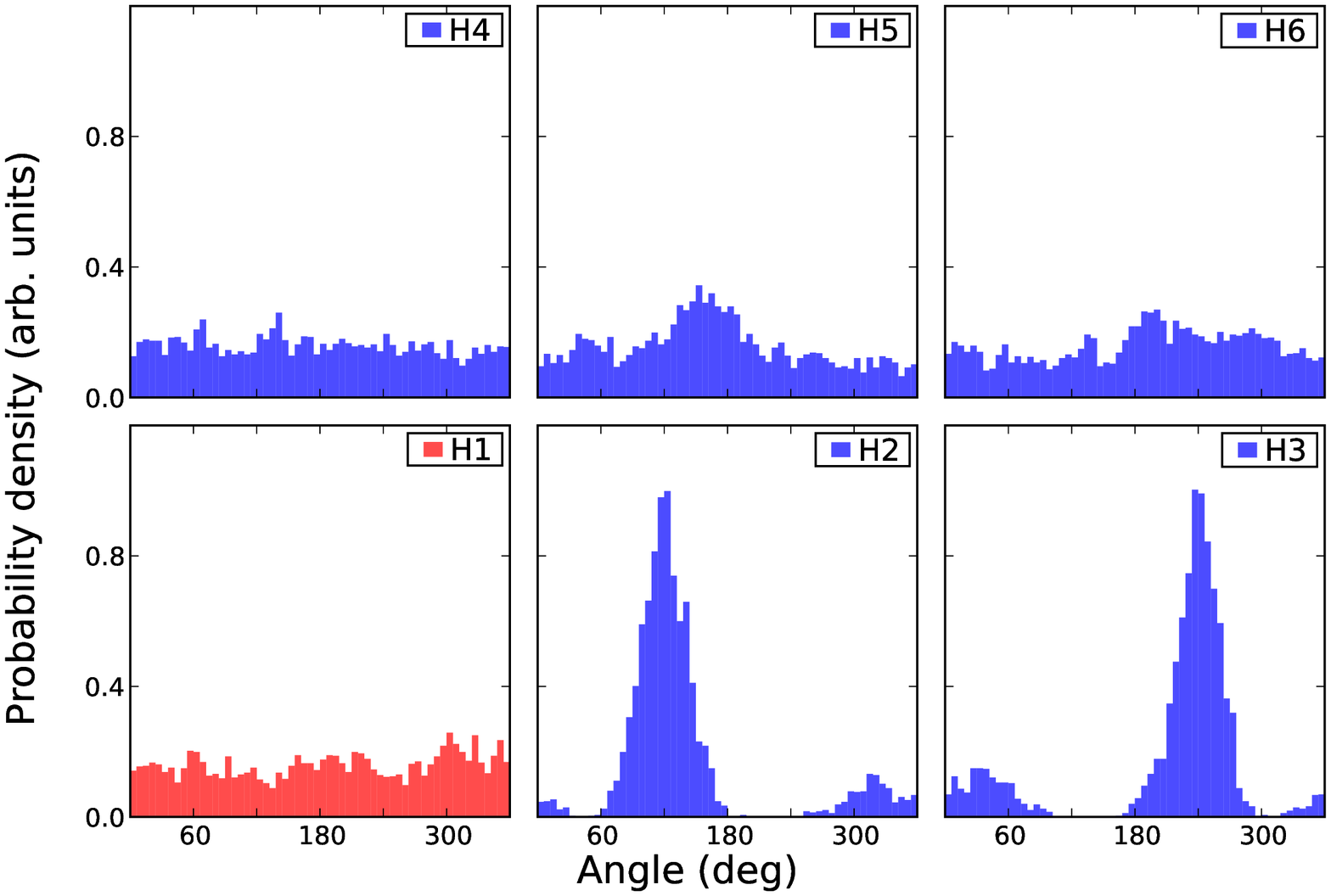}
\end{center}
\caption{\label{fig:corh} Relative angular hydrogen distributionin
P$\overline{3}$ unit cell. The H$_1$ box shows absolute angular
distribution of the H$_1$ atom. Top panels show results obtained
from the MD run at 244 K, and bottom ones at 46 K. Positions of
hydrogen atoms with labels H1,H2,$\cdots$,H6 are shown in figure
1.}
\end{figure}

To resolve this dilemma we have developed a method to quantify
angular correlations of hydrogen atoms. We extracted the
correlation between positions of hydrogen atoms with respect to
each other as well as with respect to the fixed reference crystal
lattice. The results are presented in figure~\ref{fig:corlatt} for
absolute hydrogen angular distribution and in
figure~\ref{fig:corh} for relative hydrogen angular correlations.
All these runs have been performed in the
$\sqrt{3}\times\sqrt{3}\times{}1$ unit cell. The angle $\varphi$
is an angular position of the hydrogen atom around the three-fold
axis, measured starting either from the direction of the crystal
vector ${\vec A}$ for absolute angular distributions (red
histograms), or from the direction selected by an arbitrary
reference hydrogen atom (H$_1$ [$\frac{1}{3}$,$\frac{1}{3}$,z$_1$]
in our case) for relative angular distributions (blue histograms).

The high peaks seen in figure~\ref{fig:corlatt} suggest that there
may be a lock-in of H2 and H5 hydrogen atom positions with respect
to the crystal lattice.
%
The lattice lock-in appears to be
still present at 244 K (upper panel in figure~\ref{fig:corlatt})
but we have to mention fairly strong and uniform background of
other positions in the histogram. In fact, the most of the area in
these histograms is \emph{outside} the peaks. Thus, hydrogen atoms
do not spend most of their time in the vicinity of their high
symmetry positions. This means that, despite distinct peaks
visible in the histograms, the angular positions of hydrogen atoms
are \emph{not} locked with respect to the crystal lattice
orientation.
Some of our results obtained for even lower temperatures
($T\approx 10$ K) indicate that the true lock-in with the crystal
lattice may appear at or below $T=15$ K --- at this temperature we
have observed Gaussian-shaped peaks and diminished background in
the histograms (not shown). Unfortunately, due to the difficulties
of the MD calculations performed at low temperature we were unable
to collect enough evidence to confirm this indication.

\begin{figure}
\begin{center}
\includegraphics[width=0.8\columnwidth]{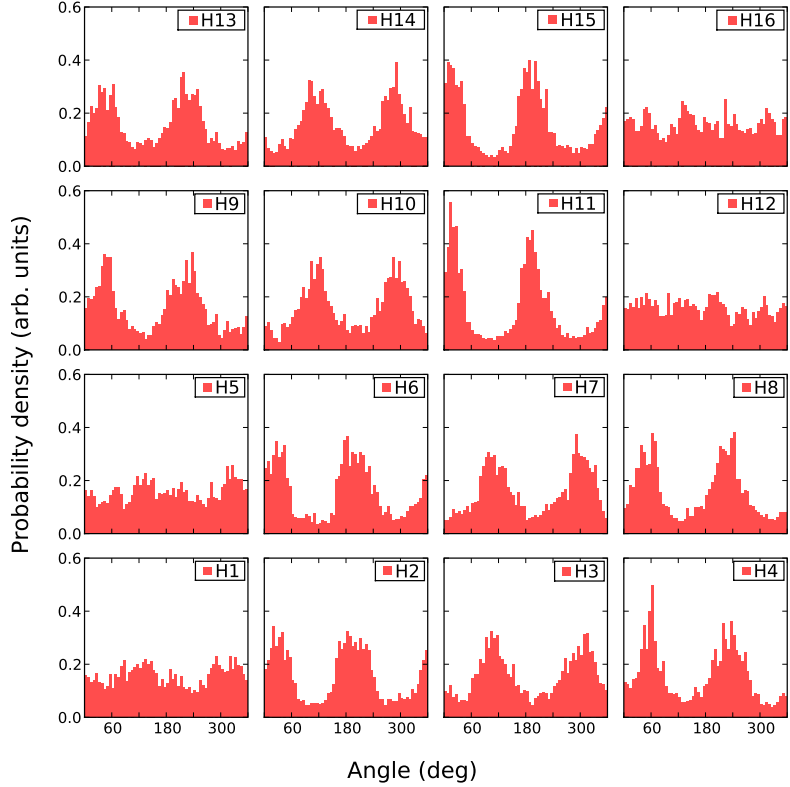}
\end{center}
\caption{\label{fig:corhl16K} Absolute angular distribution of
hydrogen atoms in the 2$\times$2$\times$2 supercell. The results
are obtained from a 6~ps MD run at 16~K. Positions of hydrogen
atoms with labels H1,H2,$\cdots$,H16 are shown in figure 1.}
\end{figure}

The relative angular distribution results displayed in figure
\ref{fig:corh} show quite a different picture. On the one hand,
there is a strong correlation, indicated by distinct peaks, of
angular position between hydrogen atoms in the same layer (H$_1$,
H$_2$,  and H$_3$ positions are as in figure
\ref{fig:structureP-3}). On the other hand, no visible
correlations were found for the atoms in the next layer (H$_4$,
H$_5$, and H$_6$). This means that the hydrogen atoms maintain
relative orientations shown in figure~\ref{fig:structureP-3},
keeping their angular positions 120$^\circ$ apart --- similar to
the angular positions in the XGT/XLT structures mentioned in
section \ref{sec:intro} --- but only if we consider atoms which
belong to the same layer of the structure. On the contrary, the
neighbouring layer positions appear only loosely coupled. These
relations are maintained up to the highest temperature we have
investigated (300 K), but broadening of the peaks observed at 244
K (upper panel of figure \ref{fig:corh}) suggests that the peaks
may disappear at higher temperatures, as one would expect.

The above result supports, to some extent, the conclusions of the
previous work on brucite crystal
\cite{Parise94,Catti95,Mookherjee06,Partin94,Parise98,Parise99,Desgranges96},
except that the hydrogen-hydrogen correlations do not appear for
all atoms, but only for neighbouring atoms in the same layer.
Furthermore, the lock-in of hydrogen positions to the lattice is
not well confirmed by the above results and probably vanishes
above relatively low temperature of $T=15$ K. These discrepancies
motivated us to investigate the system further under less
constraining conditions.

\begin{figure}
\begin{center}
\includegraphics[width=0.8\columnwidth]{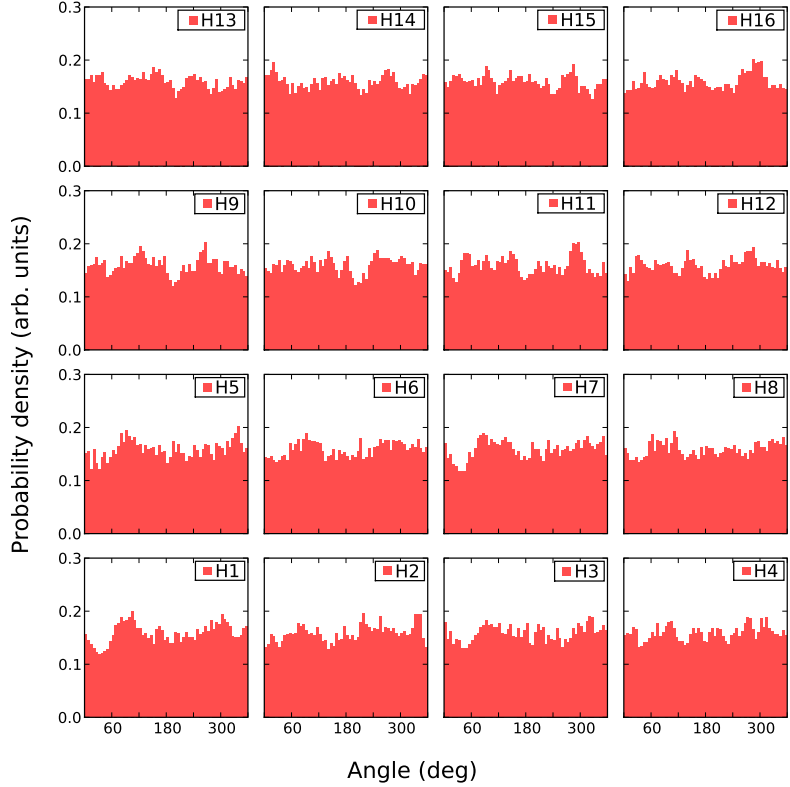}
\end{center}
\caption{\label{fig:corhl30Kabs} Absolute angular distribution of
hydrogen atoms in the $2\times 2\times 2$ supercell. The result
are obtained from a 39~ps MD run at 30~K. Positions of hydrogen
atoms with labels H1,H2,$\cdots$,H16 are shown in figure 1.}
\end{figure}

The results of the calculation performed for the larger
2$\times$2$\times$2 unit cell at temperature $T=16$ K are
presented in figure~\ref{fig:corhl16K}. It is quite clear that the
correlations seen in the smaller unit cell are changed by the
increase of the unit cell. We were unable to identify clearly any
lock-in effect of the hydrogen positions with respect to the
crystal lattice for the temperatures above 30 K
(figure~\ref{fig:corhl30Kabs}) but below this temperature the
signal arising from locked-in hydrogen positions is distinct and
robust (figure~\ref{fig:corhl16K}). We have performed the
calculations for several temperatures in the low temperature
regime and for different starting conditions and obtained the same
result in all cases.

\begin{figure}
\begin{center}
\includegraphics[width=0.8\columnwidth]{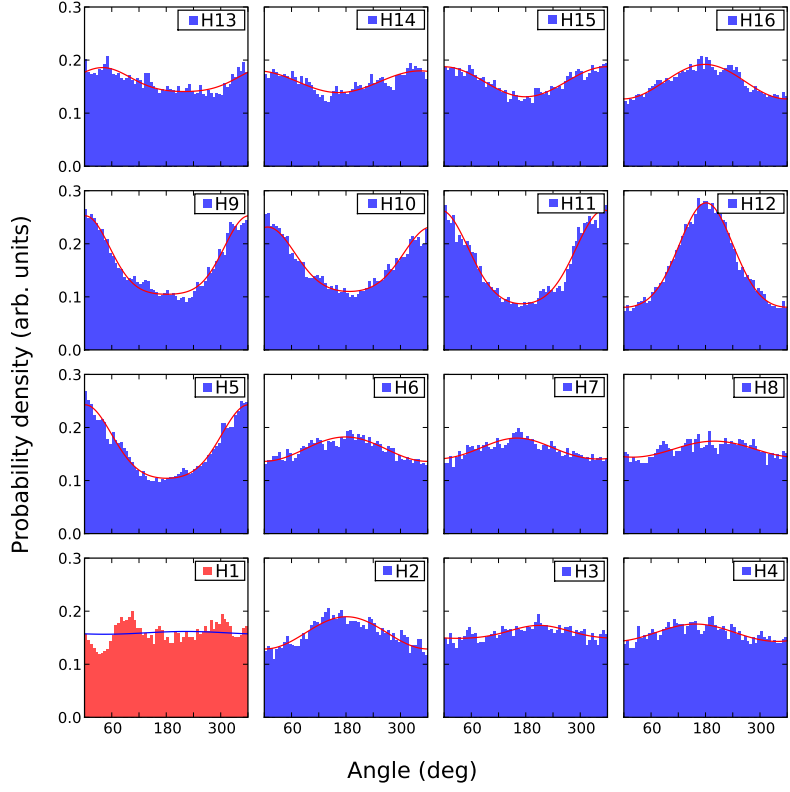}
\end{center}
\caption{\label{fig:corhl30Krel} Relative angular distribution of
hydrogen atoms in the 2$\times$2$\times$2 supercell. The results
are obtained from a 39~ps MD run at 30~K. Solid line shows the
function ${\cal F}_{\sigma}(\varphi)$ given by equation
(\ref{eq:fitfun}) fitted to the data. Positions of hydrogen atoms
with labels H1,H2,$\cdots$,H16 are shown in figure 1.}
\end{figure}

The relative angular correlations are still present when a larger
unit cell is investigated, but they are also modified with respect
to the ones found before with the smaller unit cell. The
histograms seen in figure~\ref{fig:corhl30Kabs} show distributions
of angular positions of hydrogen atoms relative to the crystal
lattice and those shown in figure~\ref{fig:corhl30Krel} ---
relative to an arbitrarily selected reference atom H$_1$, see
figure~\ref{fig:structure}. For the low temperature run
(figure~\ref{fig:corhl16K}) we can easily identify obvious
non-uniformities in the distribution of absolute angular positions
of most hydrogen atoms. The absence of distinct peaks for some
atoms may be interpreted as a result of the relatively short
simulation time (6 ps) and insufficient statistics due to the
difficulty of performing adequate sampling of configuration space
at low temperatures. We have verified that the overall picture
remains in fact similar for different runs, but individual peaks
disappear for different atoms.

\begin{figure}
\begin{center}
\includegraphics[width=0.8\columnwidth]{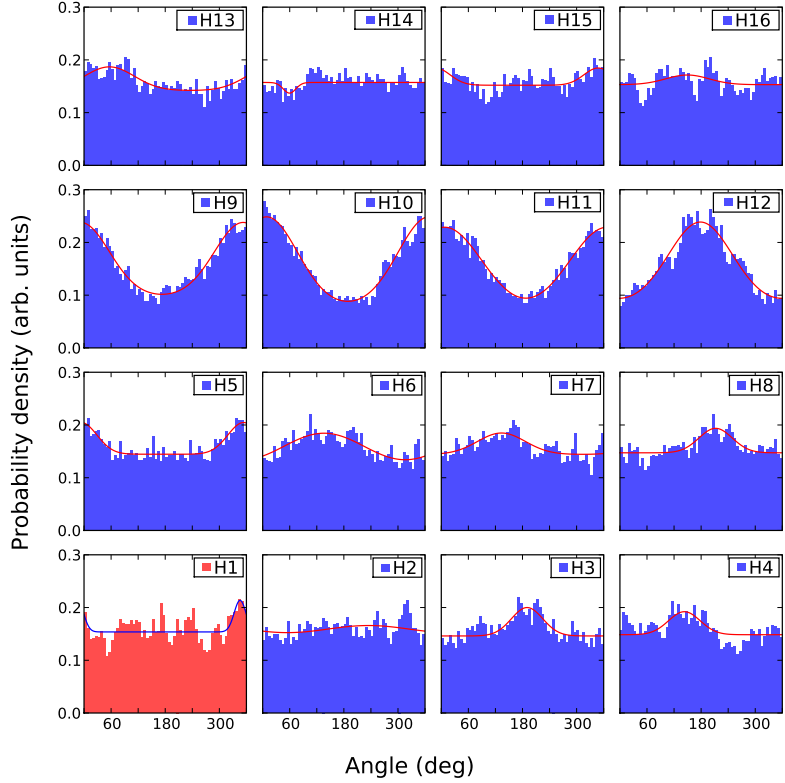}
\end{center}
\caption{\label{fig:corhl150Krel} Relative angular distribution of
hydrogen atoms in the 2$\times$2$\times$2 supercell. The results
are obtained from a 18~ps MD run at 150~K. Solid line shows the
function ${\cal F}_{\sigma}(\varphi)$ given by equation
(\ref{eq:fitfun}) fitted to the data. Positions of hydrogen atoms
with labels H1,H2,$\cdots$,H16 are shown in figure 1.}
\end{figure}

For higher temperatures, above 30-40 K, the absolute angular
positions of hydrogen atoms are completely decoupled from the
directions of crystal lattice (figure~\ref{fig:corhl30Kabs}) but
we can still observe fairly strong correlations in relative
angular positions of hydrogen atoms --- these correlations are
robust and persist even in the regime of rather high temperature.
Figure \ref{fig:corhl150Krel}, obtained for temperature $T=150$ K,
presents an essentially unchanged picture of the distribution of
hydrogen atoms' relative positions. One can notice that weaker
correlations with atoms from the first and from the last row seem
to have disappeared. However, this may be due to the smaller
statistics of the sample (18 ps vs. 39 ps) and higher noise in the
data obtained for this case.

\subsection{Angular distribution of hydrogen positions}
\label{sec:model}

We have constructed a simple probabilistic model to describe
measured angular distribution of hydrogen atoms. As a starting
point it is reasonable to assume a Gaussian distribution around
one location accompanied by a constant background filling for the
rest of the histogram. In periodic angular coordinates the
standard Gaussian distribution in angle $\varphi$ centred at $\mu$
with width $\sigma$,
\begin{equation}
{\cal P}_{\sigma}(\varphi)=
\frac{1}{\sigma\sqrt{2\pi}}\exp\left[-\,\frac{(\varphi-\mu)^2}{2\sigma^2}\right]\,,
\label{eq:gauss}
\end{equation}
should be summed up over periodic repetitions of itself, which
leads to
\begin{equation}
\frac{1}{\sigma \sqrt{2\pi}} \sum_{k=-\infty}^{\infty}
\exp\left[{-\frac{(\varphi-\mu+2\pi k)^2}{2\sigma ^2}}\right] =
\frac{1}{2\pi}\,
\vartheta_3\left(\frac{\varphi-\mu}{2},\,e^{-\frac{\sigma^2}{2}}\right)\,.
\label{eq:periodicgauss}
\end{equation}
Here $\vartheta_3$ is an elliptic Jacobi function of the third
kind. Thus, the angular distribution we introduce here reads:
\begin{equation}
{\cal F}_{\sigma}(\varphi)=
{\cal N}\left[(1-\alpha)\,\frac{1}{2 \pi}\,
\vartheta_3\left(\frac{\varphi-\mu}{2},e^{-\frac{\sigma ^2}{2}}\right)
+ \alpha\right]\,,
\label{eq:fitfun}
\end{equation}
where ${\cal N}$ is a normalizing factor. It consists of the
correlated part given by equation (\ref{eq:periodicgauss}) with
intensity $(1-\alpha)$ and the background with intensity $\alpha$.

One can notice that in the case of 30 K
(figure~\ref{fig:corhl30Krel}) as well as for 150 K
(figure~\ref{fig:corhl150Krel}) the model for the angular
distribution given by equation (\ref{eq:fitfun}) fits the obtained
numerical data remarkably well
--- it indicates the existence of a single potential minimum in
relative motion of individual hydrogen atoms. This demonstrates
that relative angular distribution reflects the correlated
behavior of the hydrogen atoms, with a single optimal angular position for
each hydrogen atom. This is obviously not the case in the absolute
angular positions depicted in figure \ref{fig:corhl16K}, where a
two-peaked distribution would be required to describe the data.

We should also note that the atoms $\{{\rm H}_9,{\rm H}_{10},{\rm
H}_{11},{\rm H}_{12}\}$ are characterized by strongly correlated
movement with respect to the atom H$_1$. They are located in the
same structure layer as the reference (H$_1$) atom. Also the
position of a single out-of-plane atom H$_5$, located directly
above the reference atom H$_1$ on the same three-fold axis, is
strongly correlated with that of atom H$_1$. Moreover, it is clear
that the directions of the peaks do not strongly follow 60$^\circ$
grid present in the smaller unit cell. The strongest signal
indicates same/opposite direction arrangement of the direct
neighbour (atom H$_9$) of the reference atom and its images in the
neighbouring unit cells (H$_{10}$-H$_{12}$). However, if we look
closer at the distributions presented in figure
\ref{fig:corhl30Krel}, we can notice small preference for image
atoms (H$_{2}$-H$_{4}$) to cluster around 120$^\circ$ and
240$^\circ$ directions. This result does not offer strong support
for the XGT/XLT arrangement hypothesis but the preferred
directions are compatible with this type of arrangement.

In light of these results we conclude that the angular
correlations in the atomic distribution obtained for the small
unit cell is likely just an artefact of the constraints imposed on
the system by the $\sqrt{3}\times\sqrt{3}\times{}1$ unit cell. We
also believe that it is still possible that the real physical
system exhibits one of the symmetries proposed before
\cite{Parise94,Catti95,Mookherjee06,Megaw73,Parise98,Parise99},
but we emphasize that it does not show these symmetries for system
size up to the $2\times 2\times 2$ supercell.

\section{Summary and Conclusions}
\label{sec:summa}

We have performed extensive calculations to determine the
structure and elastic properties of magnesium hydroxide. The
static calculations confirm the main result of the previous
research that static average structure with hydrogen atoms on the
three-fold axis is absent in the brucite crystal. Further static
calculations of the elastic constants of brucite also support
conclusions from previous papers
\cite{Parise94,Catti95,Mookherjee06,Parise98,Parise99,Desgranges96}
suggesting existence of the superstructure, probably with the
$P\overline{3}$ symmetry, and XGT arrangement of the hydrogen
atoms \cite{Mookherjee06}. This, however, does not exclude all
other possibilities which would be still consistent with the
available experimental evidence.

In the present paper we have investigated one of such
possibilities --- existence of hydrogen sublattice possibly almost
decoupled from the rest of the crystal. The results from the MD
calculations in the small supercell support this hypothesis.
Furthermore, the calculations performed in the larger
2$\times$2$\times$2 cell revealed correlations between
neighbouring hydrogen atoms at temperature between 30 K and 150 K.
We have not found any lock-in behaviour of the hydrogen system
above $T=30$ K, while below this temperature we have identified
strong correlations between angular positions of hydrogen atoms
and orientation of the lattice vectors. This result indicates that
the rotational (around the three-fold axis) degree of freedom of
hydrogen system gets decoupled from the crystal lattice around
temperature $T=30$ K which may be considered as a characteristic
energy scale for the onset of hydrogen-lattice coupling.

The 2$\times$2$\times$2 supercell (shown in figure
\ref{fig:structure}) is the largest one available for practical
use at the moment, and additional calculations of lattice dynamics
performed for the 3$\times$3$\times$3 supercell indicate that the
long-range interactions do not modify lattice dynamics in any
significant way. Thus, we expect that the main qualitative
conclusions on the crystal symmetry and hydrogen angular
correlations either would not change, or would even strengthen in
a larger system, when technology and algorithms will develop to
the point that larger systems could be investigated.

The results presented above demonstrate that the angular
coordinates of the hydrogen atoms in the brucite crystal are
likely decoupled from the crystal lattice above the estimated
characteristic temperature $T\simeq 30$ K. We have provided
evidence that, in spite of this partial disorder in the hydrogen
subsystem, angular correlations between positions of hydrogen
atoms are robust and survive also in the high temperature regime
$T\simeq 150$ K. We believe that these results will stimulate
further research and the issue of inter-hydrogen correlations in
brucite and related crystals should be further investigated using
theoretical as well as experimental methods. We suggest that
nuclear magnetic resonance spectroscopy could be a useful
experimental tool in this context.

\ack This work was partially supported by Marie Curie Research and
Training Network under Contract No. MRTN-CT-2006-035957 (c2c) and
Polish Ministry of Science and Education Grant No. 541/6.PR
UE/2008/7. A M Ole\'s acknowledges support by the Foundation for
Polish Science (FNP).

\section*{References}
\bibliographystyle{iopart-num-nourl}
\bibliography{bibliography}

\providecommand{\newblock}{}
\begin{thebibliography}{10}
\expandafter\ifx\csname url\endcsname\relax
  \def\url#1{{\tt #1}}\fi
\expandafter\ifx\csname urlprefix\endcsname\relax\def\urlprefix{URL }\fi
\providecommand{\eprint}[2][]{\url{#2}}

\bibitem{Finger89}
Finger L~W and Prewitt C~T {1989} {\em {Geophysical Research Letters}\/} {\bf
  {16}} 1395--1397

\bibitem{Kanzaki91}
Kanzaki M {1991} {\em {Physics of The Earth and Planetary Interiors}\/} {\bf
  {66}} 307

\bibitem{Kruger89}
Kruger M~B {1989} {\em {Journal of Chemical Physics}\/} {\bf {91}} 5910

\bibitem{Duffy91}
Duffy T~S, Ahrens T~J and Lange M~A {1991} {\em {Journal of Geophysical
  Research}\/} {\bf {96}} 14319

\bibitem{Redfern92}
Redfern S~A~T and Wood B~J {1992} {\em {American Mineralogist}\/} {\bf {77}}
  1129--1132

\bibitem{Fei93}
Fei Y and Mao H~K {1993} {\em {Journal of Geophysical Research}\/} {\bf {98}}
  11875

\bibitem{Catti95}
Catti M, Ferraris G, Hull S and Pavese A {1995} {\em {Physics and Chemistry of
  Minerals}\/} {\bf {22}} 200--206

\bibitem{Duffy95}
Duffy T~S, Meade C, Fei Y, Mao H~K and Hemley R~J {1995} {\em {American
  Mineralogist}\/} {\bf {80}} 222--230

\bibitem{Parise94}
Parise J~B, Leinenweber K, Weidnner D~J, Tan K and Von~Dreele R~B {1994} {\em
  {American Mineralogist}\/} {\bf {79}} 193--196

\bibitem{Sherman91}
Sherman D~M {1991} {\em {American Mineralogist}\/} {\bf {76}} 1769--1772

\bibitem{DArco93}
D'Arco P, Caus{\`a} M, Roetti C and Silvi B {1993} {\em {Phys. Rev. B}\/} {\bf
  {47}} 3522--3529

\bibitem{Raugei99}
Raugei S, Silvestrelli P~L and Parrinello M {1999} {\em {Phys. Rev. Lett.}\/}
  {\bf {83}} 2222--2225

\bibitem{Mookherjee06}
Mookherjee M and Stixrude L {2006} {\em {American Mineralogist}\/} {\bf {91}}
  127--134

\bibitem{Zigan67}
Zigan F and Rothbauer R {1967} {\em {Neues Jahrbuch fur Mineralogie
  Monatshefte}\/}  137--143

\bibitem{Partin94}
Partin D~E, O'Keefe M and Von~Dreele R~B {1994} {\em {Journal of Applied
  Crystallography}\/} {\bf {27}} 581--584

\bibitem{Megaw73}
Megaw H {1973} {\em {Crystal Structures: A Working Approach. }\/} ({W.B.
  Sanders Company, Philadelphia.})

\bibitem{Parise98}
Parise J~B, Theroux B, Li R, Loveday J~S, Marshall W~G and Klotz S {1998} {\em
  {Physics and Chemistry of Minerals}\/} {\bf {25}} 130--137

\bibitem{Parise99}
Parise J~B, Loveday J~S, Nelmes R~J and Kagi H {1999} {\em {Phys. Rev.
  Lett.}\/} {\bf {83}} 328--331

\bibitem{Desgranges96}
Desgranges L, Calvarin G and Chevrier G {1996} {\em {Acta Crystallographica
  Section B: Structural Science}\/} {\bf {52}} 82

\bibitem{Kresse96}
Kresse G and Furthmüller J {1996} {\em {Computational Materials Science}\/}
  {\bf {6}} 15--50 ISSN {0927-0256}

\bibitem{Kresse99}
Kresse G and Joubert D {1999} {\em {Phys. Rev. B}\/} {\bf {59}} 1758--1775

\bibitem{PBE96}
Perdew J~P, Burke K and Ernzerhof M {1997} {\em {Phys. Rev. Lett.}\/} {\bf
  {78}} 1396

\bibitem{Blochl94}
Blöchl P~E {1994} {\em {Phys. Rev. B}\/} {\bf {50}} 17953--17979

\bibitem{Monkhorst76}
Monkhorst H~J and Pack J~D {1976} {\em {Phys. Rev. B}\/} {\bf {13}} 5188--5192

\bibitem{Jochym99}
Jochym P, Parlinski K and Sternik M {1999} {\em {European Physical Journal B:
  Condensed Matter and Complex Systems}\/} {\bf {10}} 9--13

\bibitem{Jochym00}
Jochym P and Parlinski K {2000} {\em {European Physical Journal B: Condensed
  Matter and Complex Systems}\/} {\bf {15}} 265--268

\bibitem{Nose84}
Nosé S {1984} {\em {Journal of Computational Physics}\/} {\bf {81}} 511--519

\bibitem{Parlinski97}
Parlinski K, Li Z~Q and Kawazoe Y {1997} {\em {Phys. Rev. Lett.}\/} {\bf {78}}
  4063--4066

\bibitem{Phonon}
Parlinski K {2005} {PHONON, lattice dynamics software}

\bibitem{Jiang06}
Jiang F, Speziale S and Duffy T~S {2006} {\em {American Mineralogist}\/} {\bf
  {91}} 1893--1900

\end{thebibliography}

\end{document}